# INFLUENCE OF MATERIAL PARAMETERS ON THE PERFORMANCE OF NIOBIUM BASED SUPERCONDUCTING RF CAVITIES

Arup Ratan Jana[1,2], Abhay Kumar[3], Vinit Kumar[1,2] and Sindhunil Barman Roy[1,2]

[1]Homi Bhabha National Institute, Mumbai 400094, India
[2]Materials and Advanced Accelerator Sciences Division, Raja Ramanna Centre for Advanced Technology, Indore - 452013, India
[3]Design and Manufacturing Technology Division, Raja Ramanna Centre for Advanced Technology, Indore - 452013, India

*Abstract*

A detailed thermal analysis of a Niobium (Nb) based superconducting radio frequency (SRF) cavity in a liquid helium bath is presented, taking into account the temperature and magnetic field dependence of the surface resistance and thermal conductivity in the superconducting state of the starting Nb material (for SRF cavity fabrication) with different impurity levels. The drop in SRF cavity quality factor ($Q_0$) in the high acceleration gradient regime (before ultimate breakdown of the SRF cavity) is studied in details. It is argued that the high field $Q_0$-drop in SRF cavity is considerably influenced by the intrinsic material parameters such as electrical conductivity, and thermal diffusivity. The detail analysis also shows that the current specification on the purity of niobium material for SRF cavity fabrication is somewhat over specified. Niobium material with a relatively low purity can very well serve the purpose for the accelerators dedicated for spallation neutron source (SNS) or accelerator driven sub-critical system (ADSS) applications, where the required accelerating gradient is typically up to 20 MV/m,. This information will have important implication towards the cost reduction of superconducting technology based particle accelerators for various applications.

I. INTRODUCTION

One of the remarkable developments in the area of particle accelerators in modern times has been the successful use of the state-of-the-art superconducting radio frequency (SRF) cavities in building high energy linear accelerators (linacs) [1-5]. The phenomenon of superconductivity leads to dissipation less flow of electric current at DC level. However, when ac surface currents flow at the surface of an SRF cavity to create the required electromagnetic field for accelerating the charged particles, there is a relatively small but finite dissipation of heat, which increases with the frequency ($f$) of ac surface current as $f^2$ [1, 5]. Even then for generating a given value of acceleration gradient $E_{acc}$, the power loss $P_c$ at the cavity surface in an SRF cavity is significantly small as compared to that in the case of a normal conducting radiofrequency (RF) cavity, when the operating frequency is in the range below ~ 3 GHz [1]. Therefore, the SRF cavities are quite attractive choice for high energy - high current accelerators, operating in the continuous wave (CW) or long pulse mode [5, 6]. The low loss feature of an SRF cavity is characterized by its extraordinary high value of quality factor $Q_0$ (~ $10^{10}$), which is inversely proportional to the power loss $P_c$ at the cavity wall [1, 5, 7]. The superconducting material used for making the SRF cavity is characterized by its surface resistance $R_s$ in the superconducting state at the operating frequency. The power loss of an SRF cavity is proportional to $R_s$, which implies that $Q_0$ will be inversely proportional to $R_s$ [5, 7].

Niobium (Nb) is the material of choice for making SRF cavities because of its relatively high value of superconducting transition temperature or critical temperature $T_c$ (~ 9.2 K) as well as the lower critical magnetic field $B_{c1}$, relative abundance and ease in availability, and mechanical strength as well as formability. Experimentally, the $Q_0$ of a Nb-SRF cavity shows the following typical trend with the increasing strength of the amplitude $B_a$ of the RF magnetic field at the cavity surface: it first increases slightly in the very low field ($B_a$~0 to 20 mT), then it decreases gradually in the medium field regime ($B_a$~20 to 80 mT), and finally, a sharp fall occurs at higher RF fields ($B_a$~80 to 180 mT), which is known as the $Q_0$ drop [8, 9]. This sharp fall in $Q_0$ indicates the breakdown of the superconductivity in the SRF cavity material. The corresponding value of $B_a$ at which this happens is known as the threshold magnetic field $B_{th}$. Theoretically, the limit of the performance of an Nb-SRF cavity is reached when the oscillatory magnetic field (rather magnetic flux-lines) associated with the applied RF field starts penetrating the bulk of Nb material giving rise to the heat dissipation. This is expected to happen at the lower critical field ($B_{c1}$) [5,10,11]. In some quarters, it is believed that this dissipation-less superconducting response may continue beyond $B_{c1}$ up to the superheating field $B_{sh}$ [5]. However, it has been experimentally observed that in Nb-SRF cavities, $B_{th}$ is often significantly less than the lower critical field $B_{c1}$ [5]. In recent times, there is a continual quest in the SRF community to push this threshold limit $B_{th}$ towards $B_{c1}$ (or beyond) of Nb to achieve a higher value of accelerating gradient $E_{acc}$, and simultaneously a higher value of $Q_0$ to make the higher energy accelerators economically more viable.

More importantly, the observed threshold value $B_{th}$ of Nb-SRF cavities depends on the quality of the starting Nb material, as well as the processing techniques used during the cavity development. The high purity of the Nb material ensures a higher value of the thermal conductivity $\kappa$ in the normal state and the cavity processing removes the surface damage of the Nb material, which takes place in the course of forming a SRF cavity. However, at the typical operating temperature of 2K in the superconducting state of Nb, the value of $\kappa$ reduces significantly from it its value in the normal state just above $T_C$ [1, 5]. Therefore, the heat removal turns out to be a crucial issue even though the rate of heat generation may be small in the case of an SRF cavity.

In order to realize the goal of high accelerating gradient accompanied with high $Q_0$, the prevalent practice followed in the SRF community is to use highly pure niobium, mainly to achieve a higher normal state thermal conductivity [1]. The purity of a metal is often characterized by the residual resistivity ratio (*RRR*), which is usually defined as the ratio of the resistances of the metal at room temperature and at a low enough temperature, where the resistance of the metal has reached its residual resistance limit [1, 5, 6]. Contemporary SRF community has set the value of *RRR* ~ 300 as the most recommended choice for the niobium material for SRF cavity fabrication. Experimental observations are there both in favor as well as against this empirical choice of standard for *RRR* [6]. For a metal/superconductor, the increasing purity level results in an improvement in the thermal conductivity in the normal state, but in a superconductor, in the clean limit, the value of superconducting surface resistance $R_s$ also *increases* with purity as predicted by the BCS theory. Nb metal with *RRR* value greater than 50 is expected to be in the clean limit of superconductivity, and the increase in $R_s$ with the increase in *RRR* value beyond 50 has actually been observed experimentally [12]. Therefore, for a high purity Nb material, although the rate of heat conduction will increase, this advantage may be countered by the fact that there will be *more* heat generation at the cavity surface. A rigorous calculation of the heat transfer problem will therefore be needed to find out the amplitude

$B_a$ of magnetic field at the surface of the SRF cavity at which the thermal breakdown of superconductivity occurs in the cavity for a given grade of purity of the starting Nb material for SRF cavity fabrication. More importantly, $R_s$ is also influenced by the magnetic field $B_a$. Hence, it is no longer only a thermal phenomenon, but a *magneto-thermal* phenomenon [10, 13]. A theoretical analysis to calculate the breakdown field has been performed in *Refs.* [14, 15], where $R_s$ is assumed to be independent of $B_a$, and is a function of temperature alone. Weingarten [16] and Gurevich [17] have taken exponential dependence of $R_s$ on $B_a$ into account, and have performed more rigorous analyses of the thermal breakdown phenomenon.

Most of these analyses of the thermal breakdown phenomena did not consider the temperature dependence of the thermal conductivity $\kappa$ of niobium in spite of the same being significant. A more rigorous approach will therefore be to include the dependence of temperature on $\kappa$, and dependencies of magnetic field as well as temperature on $R_s$ in the analysis, for *different purity levels of niobium*. In this paper, we have followed this approach to perform a theoretical analysis of this *magneto-thermal* process in a self-consistent manner, and finally calculated the steady state temperature profile inside the material of the Nb SRF-cavity. Knowing the temperature of the inner surface of the SRF cavity, the surface resistance and therefore the $Q_0$ value is evaluated as a function of $B_a$ for different purity levels of the cavity material. This methodology is then used to evaluate the threshold magnetic field $B_{th}$ as a function of the material purity level and then to find the most optimum value of the material parameters to enhance the performance of an SRF cavity.

To the best of our knowledge, such an analysis has not been performed in the past. We believe that it is important to perform this kind of analysis rather than specifying a high value of purity of the starting Nb material for SRF cavity fabrication on an ad-hoc basis.

The analysis presented in the paper is for an operating frequency of 1300 MHz, which has been chosen as the operating frequency for the Tera Electron Volt Energy Superconducting Linear Accelerator (TESLA) cavities for the proposed linear electron-positron collider. Similar type of elliptic SRF cavities with fundamental frequency of 650 MHz will also be used in the injector linac for the proposed Indian Spallation Neutron Source (ISNS) project [18, 19] as well as other projects such as Chinese-ADS program and PIP-II project. It is important to note that for the collider applications, the emphasis is more on higher accelerating gradient (typically in the range 40 - 50 MV/m), whereas in the case of spallation neutron source (SNS) or accelerator driven system (ADS) applications, beam dynamics considerations set the required value of accelerating gradient to be typically up to 20 MV/m, and it is more important here to reduce the heat loss in the cavity. It is therefore more prudent to assess whether the high purity niobium having RRR ~ 300 (which is relatively expensive compared to lower RRR grade niobium) is really necessary for SRF cavity development. The paper is organized as follows. SECTION II discusses the analytical models used to calculate the thermal conductivity $\kappa$ and superconducting surface resistance $R_S$ as a function of (*i*) the purity level of the Nb material, (*ii*) RF magnetic field amplitude $B_a$ at the cavity surface, and (*iii*) temperature $T$. Next, in SECTION III, we present the results of our magneto-thermal analysis, where we highlight the influence of the purity level of niobium on the electromagnetic response of an Nb-SRF cavity. Finally, in SECTION IV, we discuss the important inferences that can be drawn from the analysis presented in the paper, and conclude.

II. THEORETICAL FORMULATION

**II.A**. *Generalities*

The quality factor $Q_0$ of an SRF cavity is a measure of the amount of RF power $P_c$ dissipated at the cavity wall corresponding to the electromagnetic energy $U$ stored inside the cavity [8]. It is evaluated using the formula $Q_0 = G/R_s$, where $G = 2\mu_0^2 \omega U / \int_S B_a^2 \, dS$ is solely dependent on the cavity geometry, and is known as the geometry factor [5, 7]. Here, $\omega$ is $2\pi$ times the RF frequency $f$, $\mu_0$ is the permeability of free space, and the integration is carried over the inner surface area of the SRF cavity. If we assume that $R_s$ is field-independent, then $Q_0$ will have a very weak dependency on $B_a$, and should remain nearly constant up to the breakdown limit. But the experimentally observed quality factor is associated with a $Q_0$-slope. Also, the breakdown does not occur at a sharp value of $B_a$, instead it occurs over a range of $B_a$. This implies that $R_s$ should have some direct or indirect functional dependency on $B_a$ [15, 16]. This will be discussed in the next sub section.

It may be in order to present here a brief discussion on the purity level of the material. For niobium, mostly the defects are of two types – (*i*) impurities due to metallic (*e.g.*, Ta, Fe, Sn *etc.*) or non-metallic (*e.g.* O, H *etc.*) inclusions, and (*ii*) various kinds of material defects including dislocations [20]. Although the first type of defect is reduced by following an expensive processing and purification process of the niobium material, the second type of defect, *i.e.*, dislocations, is unavoidable even in very pure Nb. The amount of such defects will actually increase during the half-cell formation of an elliptical Nb-SRF cavity, and thus the *RRR* of the Nb material in a finished product of Nb-SRF cavity will be significantly different from the *RRR* of the starting Nb material. In general, the electronic mean free path ($l_e$) of a metal is a function of the purity level of that material [5,11,16,21]. The normal state electrical resistivity ($\rho_{no}$) of a metal can be estimated from the value of the mean free path $l_e$. For niobium, at $T_c \sim 9.2$ K, we can write $l_e = (3.7 \times 10^{-16} \, ohm - m^2)/\rho_{no}$ [20], following the SI units. We would like to emphasize that for the normal electrons, the value of $\rho_{no}$ as well as $l_e$ remain almost unaltered in Nb in the temperature range below $T_c$. As already mentioned, the commonly followed approach to quantify the purity level in Nb is in terms of *RRR*, which is the ratio between the resistivity $\rho_{300K}$ at 300 K and the normal state resistivity ($\rho_{no}$) at a sufficiently low temperature, say at 9.2 K, i.e., just above the superconducting transition temperature. Therefore, the *RRR* will be proportional to $l_e$, assuming that $\rho_{300K}$ is nearly independent of purity level of the material. In the next sub sections, we will explain how the level of impurity plays an important role in deciding the $R_S$ and $\kappa$ of a material.

**II.B.** *Electrical surface resistance* ($R_s$)

The surface resistance $R_s$ of a superconducting material is obtained as the sum of BCS resistance $R_{BCS}$, and residual resistance $R_i$. As it is already mentioned $R_s$ has a strong dependence on the electronic mean free path $l_e$. In the *dirty limit*, $l_e$ becomes adequately small. Therefore, in this case, RF field remains nearly constant during the time interval between two collisions. This scenario changes with the increasing level of purity of the material, and in the *clean limit*, the temporal variation of the field is noticeable during the time interval between two collisions. In order to include this *non-local* response of electromagnetic fields, we followed a procedure adopted in the computer code SRIMP [21, 22], which uses the full BCS theory in the calculation of $R_{BCS}$. Figure 1 shows the plot of $R_{BCS}$ at a constant temperature 2 K, as a function of $l_e$ at *zero-magnetic-field*, *i.e.*, $B_a = 0$. As shown in the figure, after a shallow minimum near $l_e \sim 10$ nm, the value of $R_{BCS}$ increases gradually with $l_e$ in the *clean limit* of the superconducting material.

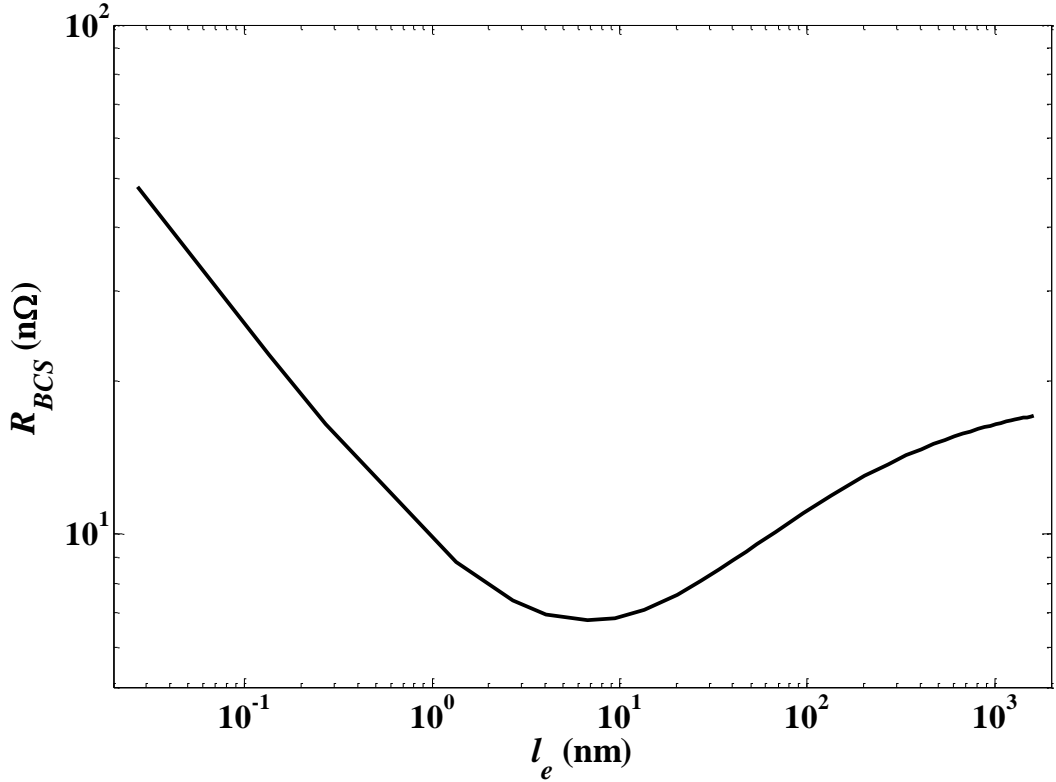

Fig.1: Plot of $R_{BCS}$ at 2 K as a function of $l_e$ for niobium.

In order to calculate $R_{BCS}$ using this formulation, we considered the *zero-temperature* coherence length $\xi_0$ and London penetration depth $\lambda_0$ as 39 nm and 33 nm, respectively[16], and the superconducting band gap $\Delta = 1.9 k_B T_c$, where, $k_B$ is the Boltzmann constant.

In the presence of an applied magnetic field $B_a$, the expression of $R_s$ gets modified. Following the work of Gurevich [17], for a type-II superconductor in the clean limit, the modified $R_s$ ($T$, $l_e$, $B_a$) can be written as follows:

$$R_s = \frac{8 R_{BCS}(l_e)}{\pi \beta_0^2} \int_0^\pi sinh^2\left(\frac{\beta_0}{2} \cos\tau\right) tan^2\tau d\tau + R_i, \quad [1]$$

where, $\beta_0 = \frac{\pi}{2^{3/2}} \frac{B_a}{B_c} \frac{\Delta}{k_B T}$, $B_c$ is the thermodynamic critical magnetic field, which is 200 mT for Nb [17], and $R_i$ is the residual resistance, which is present even at zero temperature, and has its origin in trapped magnetic flux, formation of niobium hydride islands near the surface *etc.* [11]. Note that although the term "residual resistance" in Eq. (1) may appear similar to residual resistance in the definition of *RRR*, they are completely different and independent of each other. The term *resistance* in "residual resistance $R_i$" actually denotes the surface resistance [7]. Based on the experimentally observed values, we have assumed a value of 5 nΩ for $R_i$ in our analysis. The dependency of $R_S/R_{BCS}$ as a function of the parameter $\beta_0$ is shown in Fig. 2. A precise estimation for the surface resistance as a function of the temperature, the purity level of the superconducting Nb material, and the magnetic field can be obtained with the help of Eq. (1).

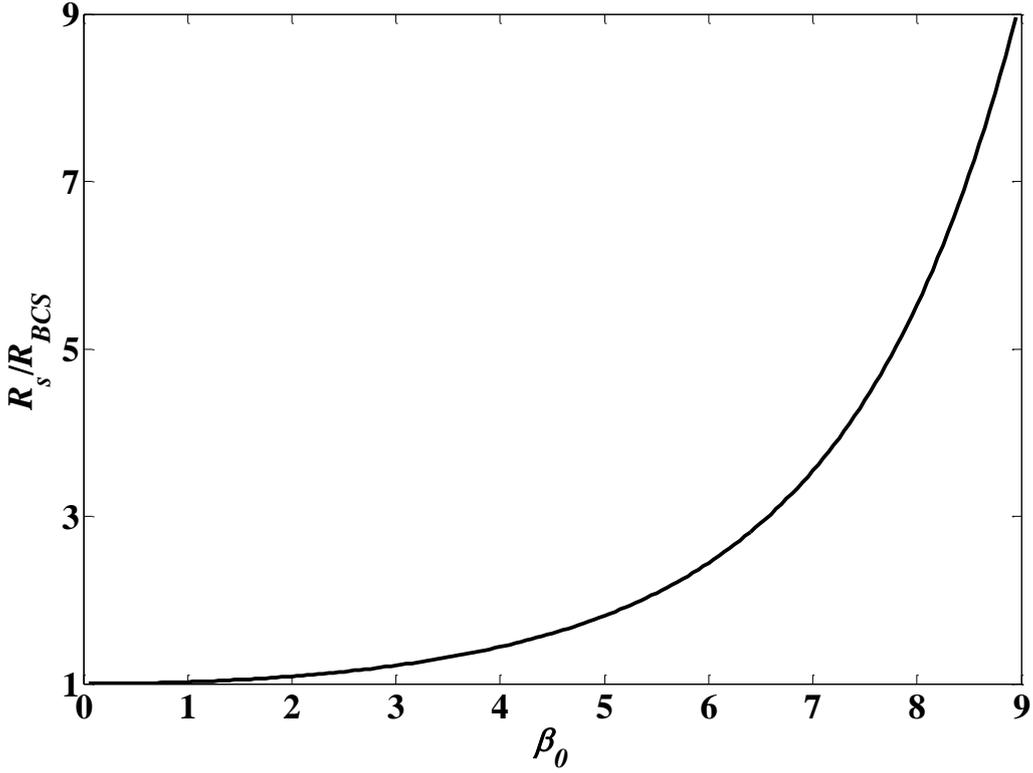

Fig. 2: Plot of normalized $R_S$ as a function of $\beta_0$.

In the high field limit, Eq. (1) can be written in an approximate form as follows:

$$R_s \cong \frac{4R_{BCS}}{\beta_0^3\sqrt{2\pi\beta_0}} e^{\beta_0} + R_i. \qquad [2]$$

This shows that in the high field, superconducting energy gap $\Delta$ will be reduced as $\Delta_{eff} = \left(1 - \pi B_a/(B_c 2^{3/2})\right)\Delta$. This modification in the energy gap will accelerate the effective breakdown phenomena.

In the next sub-section, we will discuss the dependence of thermal conductivity of niobium on different parameters, in the superconducting state.

**II.C.** *Thermal conductivity of the SRF cavity material*

There are two types of heat carriers in a metal - the conduction electrons, and the lattice vibrational modes *i.e.,* phonons [21]. Amongst these two, in typical metals the electronic contribution dominates. The total thermal conductivity of a metal $\kappa(T)$ is the summation of these two contributions, *i.e.* $\kappa(T) = \kappa_{en}(T) + \kappa_L(T)$ [24, 25]. The electronic contribution $\kappa_{en}(T)$ arises because of the scattering of normal electrons from lattice imperfections due to the thermal vibrations as well as various defects (including impurities) present in the material [25], which can be estimated using the Wiedemann-Franz law (at low temperatures), which is stated as $\kappa_{en} = L_0\sigma_{no}T$ [25], where, $L_0$ is the Lorentz number. Considering the contribution from the electron - lattice scattering *i.e.* $\kappa_{el} = 1/aT^2$, where $a$ is constant, the total electronic thermal conductivity can be written as $\kappa_{en}(T) = (1/L_0\sigma_{no}T + aT^2)^{-1}$. As discussed in the previous paragraph, with the increase in the purity level of the material, its electrical conductivity $\sigma_{no}$ increases and so does the $\kappa_{en}(T)$. Hence, the material in its purest form will offer the best thermal conductivity.

In the superconducting state of a metal, the number of free electrons reduces because of the formation of cooper pairs. This results in a scaled down contribution in the electronic thermal conductivity $\kappa_{es}(T)$ of a superconductor. This scale factor $R(y)$, as given by Bardeen-Rickayzen-Tewordt [26] is as follows:

$$\frac{\kappa_{es}}{\kappa_{en}} = R(y) = \frac{1}{f(0)}\left[f(-y) + y\ln(1 + \exp(-y)) + \frac{y^2}{2(1 + \exp(-y))}\right], \quad [3]$$

where $f(-y)$ is the Fermi integral, and is defined as $f(-y) = \int_0^\infty (z/[1 + \exp(z + y)]dz$, and $y = \Delta(T)/2k_B T$. Figure 3 shows a plot of $R(y)$ as a function of temperature.

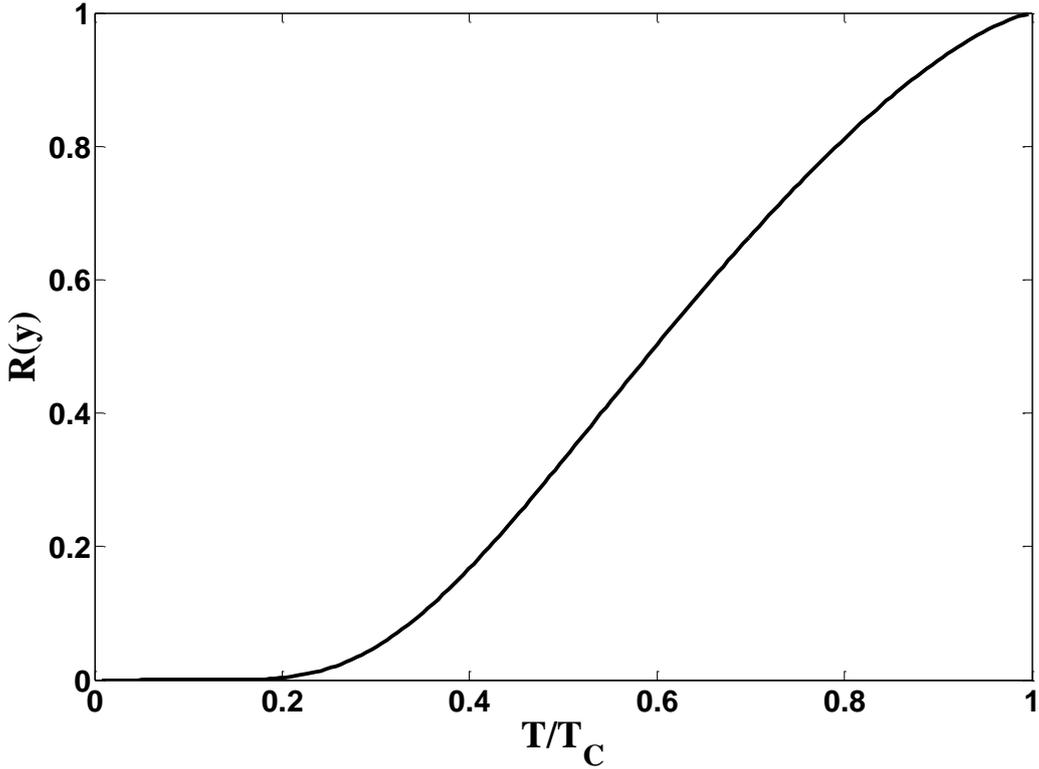

Fig. 3: Plot of *R(y)* as a function of *T/T_c*.

In our analysis, we have estimated $\kappa_{en}$ for different values of the impurity levels *i.e.*, for different values of $\sigma_{no}(l_e)$, and to calculate the normal state thermal conductivity of Nb at 9.2 K, we have used $L_0 = 2.45 \times 10^{-8}$ W K$^{-2}$ [23,24] and $a = 7.52 \times 10^{-7}$ m W$^{-1}$ K$^{-1}$ [23] respectively.

Unlike the free electrons, crystal lattice contributes in a relatively small amount in the total thermal conductivity. The total $\kappa(T)$ for a material in its superconducting state can be estimated from the following equation [24,25]:

$$\kappa(T) = \kappa_{es}(T) + \kappa_L(T) = R(y)\left(\frac{1}{L\sigma_{no}T} + aT^2\right)^{-1} + \left(\frac{1}{DT^2 e^y} + \frac{1}{Bl_{ph}T^3}\right)^{-1}. \quad [4]$$

Here, the second part in the right-hand side is the phononic contribution due to the lattice, where *D* and $Bl_{ph}$ are the two constants, and $l_{ph}$ is the phonon mean free path. The values of these two constants depend on different levels and types of post processing [5] that the cavity has undergone. For a defect-free metal with high purity, there is the likelihood of a phonon peak at a very low temperature (around $T\sim 2$ K), which can result in an enhancement in $\kappa(T)$. However, for a non-annealed SRF cavity, defects and dislocation introduced during the process of forming an SRF cavity destroys the phonon peak, partly or sometimes completely. These conditions however improve with the post-processing of an SRF cavity. Figure 4 shows the variation of thermal conductivity of niobium with temperature for three different cases. First, the case of pre-strained, small grain sample of Nb is considered, which shows a phonon peak in thermal conductivity near 2 K [27]. The second case is without the phononic contribution. Finally, the third case corresponds to a practical situation, where the phonon peak is not completely destroyed, but is scaled down suitably in accordance with the experimentally observed results at 2 K in Ref. 28.

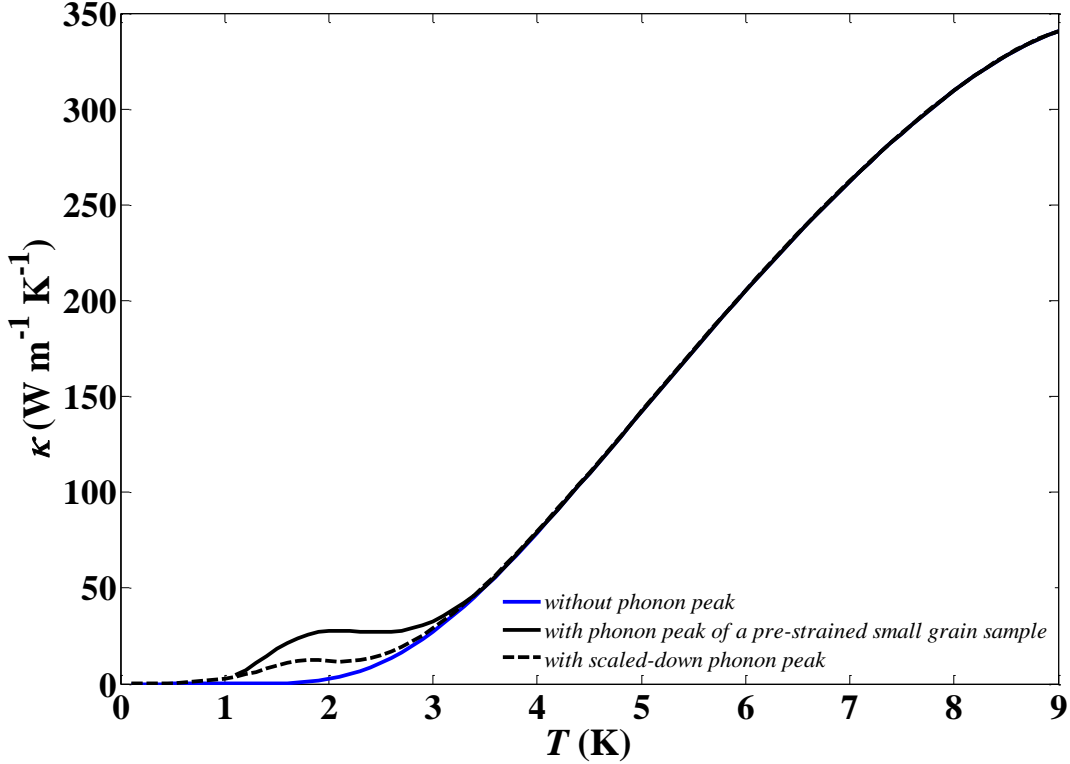

Fig. 4: Total thermal conductivity $\kappa$ as a function of the temperature $T$. The blue curve denotes the case without the phononic contribution. Enhancement in $\kappa$ at low temperature due to phonons is observed in a pre-strained, small grain Nb sample [27], as shown in the continuous black curve. The dotted black curve shows the case with reduced phonon peak in $\kappa(T)$, in accordance with experimental observation in Ref. [28] for an SRF cavity.

As it is expected, improvement in $\kappa(T)$ is more effective if we keep the liquid helium bath temperature $T_B = 2$ K. The phonon peak has almost no effect if we consider $T_B$ to be equal to 4.2 K.

The thermal conductivity of Nb is dependent on the applied RF magnetic field. However, we did not incorporate this dependency in our calculation. This is because in a superconducting cavity, the RF electric and magnetic fields almost vanish in the bulk of the material.

Next, we discuss the Kapitza resistance that is developed at Nb-He bath interface, and contributes prominently in the low temperature regime, causing a temperature jump $\Delta T = (T_S - T_B)$ across the interface, where, $T_S$ is the temperature of the cavity outer wall. The value of $\Delta T$ decides the amount of heat flow $\bar{Q}$ per unit interface area per unit time, given by $\bar{Q} = h_k(T_S - T_B)$. Here, $h_k$ is the Kapitza conductance, which is a function of $T_S$ and $T_B$. It is estimated in the unit of W m$^{-2}$ K$^{-1}$ from the following equations [29] for $T_B \sim 2$ K.

$$h_k = 200 T_S^{4.65} \left[ 1 + 1.5 \left( \frac{T_S - T_B}{T_B} \right) + \left( \frac{T_S - T_B}{T_B} \right)^2 + 0.25 \left( \frac{T_S - T_B}{T_B} \right)^3 \right] \quad [5]$$

Hence, finally in the steady state condition, the heat balance equation is written as:

$$\frac{1}{2\mu_0^2} R_S(T_{so}, B_a, l_e) B_a^2 = -\kappa(T, l_e) \nabla(T) = h_k (T_S - T_B) \quad [6]$$

Here, $T_{so}$ is the steady sate temperature of the cavity inner wall.

III. NUMERICAL CALCULATIONS AND ANALYSIS OF RESULTS

In this section, we discuss the analytical results of our magneto-thermal analysis, where the purity level of the material is considered as an input parameter. In this analysis, the inner surface of SRF cavity is the source of the outward heat flux, which is then diffused through the thickness of the wall, and is finally dissipated in the liquid helium bath maintained at a constant temperature $T_B$. The amount of heat flux depends on $R_s(T, B_a, l)$, and the rate of heat diffusion is controlled by $\kappa(T, l)$ as well as $h_k$ ($T, T_B$). The calculation of $R_s$, $\kappa$ and $h_k$ is performed using the formulation described in the previous section. We then use Eqs. (5) and (6) to find out the temperature of the cavity inner surface in the steady state. The surface resistance $R_s$ is evaluated at this temperature, including the effect of $B_a$, for the given value of $l_e$. The quality factor $Q_0$ is then calculated using this value of $R_s$.

In the remaining part of this section, we perform the calculations for a 1300 MHz SRF cavity, taking the functional dependency of $\kappa$ and $h_k$ into account. We first described the details of problem modeling, followed by presentation of results of numerical calculations in two sub-sections.

III. A. *Simulation model*:

Fig. 5 describes the model, which is a 2.8 mm thick, infinite Nb slab with planar geometry. One side of this slab is exposed to a spatially uniform RF field resonating at 1300 MHz, whereas the other side is in contact with liquid helium at a bath temperature $T_B$. From the symmetry of the problem, the heat diffusion equation will be one dimension (1D) here.

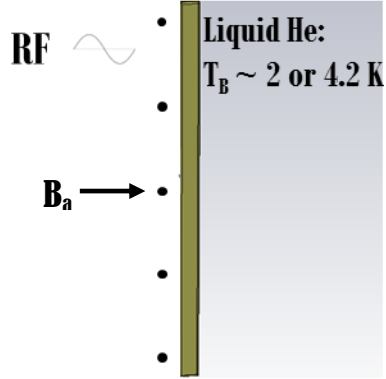

Fig. 5: Geometry of a 2.8 mm thick infinite plate used as the model. Here, the 'dot's represent the applied magnetic field $B_a$ on the surface

III.B. *Numerical calculations and results*

In order to obtain the steady state solutions for the converged values $R_s$ and $\kappa$, computer programs were written in MATLAB. As mentioned earlier, the phonon peak influences the thermal conductivity calculation more prominently near $T \sim 2$ K. Detailed *magneto-thermal* calculations were performed for all the three scenarios shown in Fig. 4. We first performed the *magneto-thermal* analysis considering a fixed value of $\sigma_{no} \sim 2.069 \times 10^9$ Ω m$^{-1}$. Using the expression for *RRR* given in *Refs*.[1] and [5], this corresponds to $RRR \sim 300$. Figure 6 shows variation of $Q_0$ as a function of the applied magnetic field $B_a$ for the geometry shown in Fig. 5.

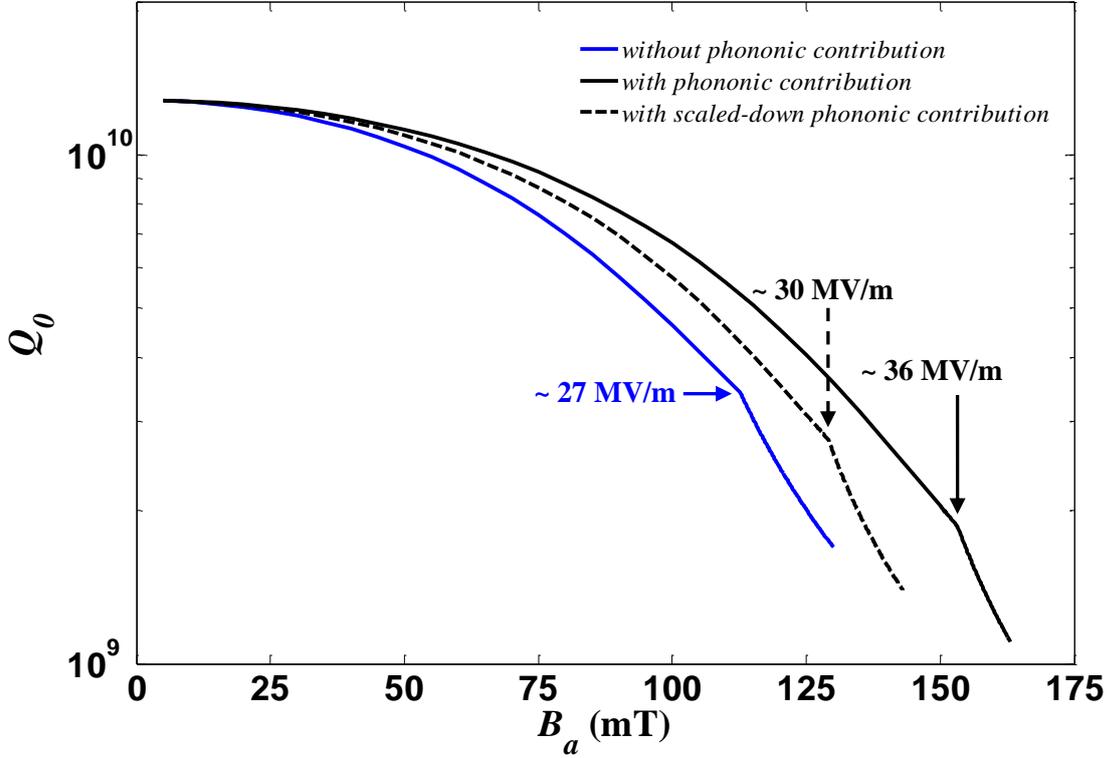

Figure 6: Plot of $Q_0$ as a function of $B_a$ for 1.3 GHz TESLA cavity, considering three possible variations of $\kappa(T)$ described in Fig. 4.

As shown in Fig. 6, the approximate value of $B_{th}$ at which there is a sharp change in the rate of decrease in $Q_0$ is 114 mT, when we do not consider the phononic contribution in $\kappa(T)$. Value of $B_{th}$ increases to 154 mT, when we consider full phononic contribution in $\kappa(T)$, and to 130 mT, when we consider a scaled down phononic contribution. Using the ratio of peak surface field $B_{pk}$ to accelerating field $E_{acc}$ specified for the optimized geometry of TESLA cavity in *Ref.* 1, we obtained the maximum achievable value of acceleration gradient $E_{acc}$ as 27 MV/m, 30 MV/m and 36 MV/m for no phononic contribution case, scaled down phononic contribution case and full phononic contribution case, respectively. We would like to point out that our theoretical prediction without the phonon peak is in good agreement with the experimentally reported observation in Fig. 12 of *Ref.* 1, where a similar trend is seen and a similar value is obtained for maximum achievable $E_{acc}$. Our result considering the contribution of full phononic contribution shows agreement with the experimental obtained value of ~40 MV/m in *Ref* 28. Reasonable agreement between the experimentally obtained results with the results of our analytical calculation benchmark the approach followed in our analysis.

The main objective of our analysis is to study the effect of the purity level of niobium on the threshold value $B_{th}$ of the RF magnetic field at the SRF cavity surface. We have described in previous sections that the mean free path $l_e$ of the normal electrons gives an estimation of the purity level of the material. However, $l_e$ is not a directly measurable parameter. Hence, we can quantify the purity of the material in terms of its normal state electrical conductivity $\sigma_{no}$, which is directly measurable. Next, we repeat the calculation for different values of $\sigma_{no}$ and obtained the threshold values of the RF magnetic field $B_{th}$ as a function of $\sigma_{no}$, which is shown in Fig. 7.

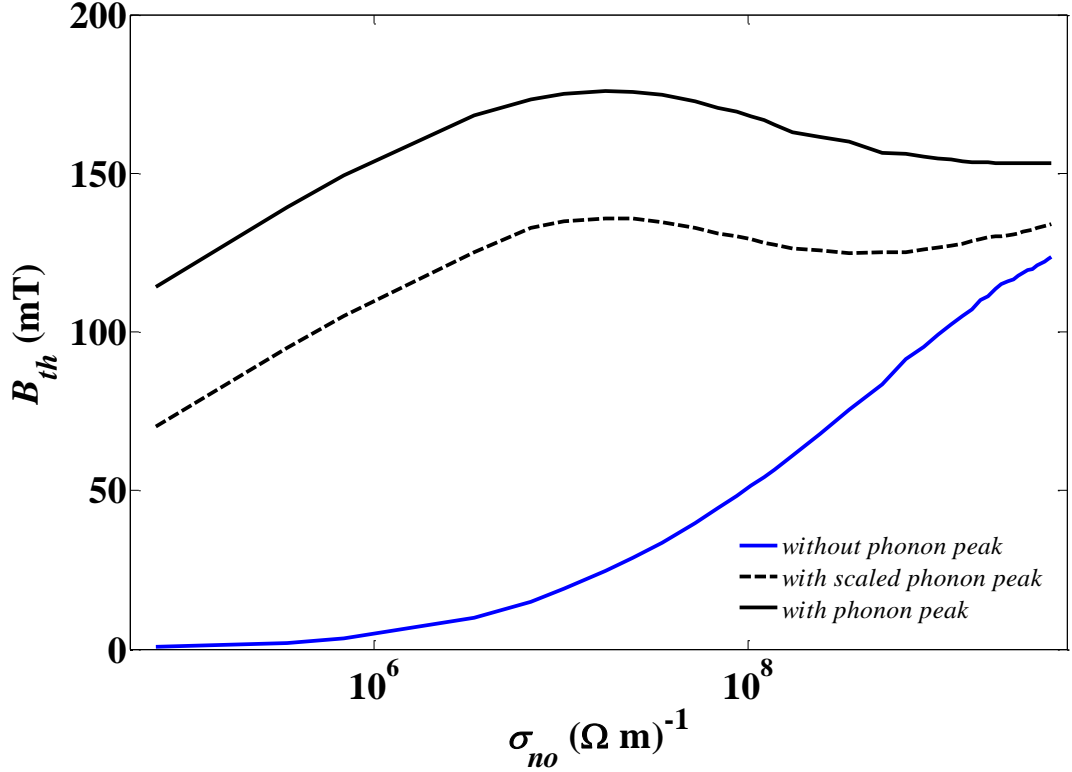

Fig. 7: Plot of $B_{th}$ as a function of $\sigma_{no}$. Here, the blue curve corresponds to the case where the phononic contribution is not considered, whereas the continuous and dotted black curves correspond to the case of full phononic contribution, and scaled down phononic contribution, respectively.

As seen in Fig. 7, for the case without phononic contribution, $B_{th}$ initially shows a rapid and monotonic rise with $\sigma_{no}$, and finally for higher value of $\sigma_{no}$ corresponding to high purity of Nb, the rate of rise decreases. Next, we discuss the case with full phononic contribution. Here, interestingly, $B_{th}$ initially increases with $\sigma_{no}$, reaches a maximum value of ~ 176 mT at $\sigma_{no} \sim 1.724 \times 10^7$ $\Omega$ m$^{-1}$, and finally for higher value of $\sigma_{no}$, $B_{th}$ saturates at ~ 153 mT. For the case with scaled down phononic contribution, $B_{th}$ reaches a maximum value of 134 mT at $\sigma_{no} \sim 1.724 \times 10^7$ $\Omega$ m$^{-1}$, and saturates at ~ 125 mT for higher values of $\sigma_{no}$. Based on these results, we can make interesting comparison between expected performances from *RRR* 300 and *RRR* 100 grade Nb cavities. For the case with phononic contribution, $B_{th}$ is nearly the same for *RRR* 300 and *RRR* 100 cases. On the other hand, for the case without phononic contribution, $B_{th}$ decreases from 114 mT for the *RRR* 300 case to 91 mT for *RRR* 100 case. We would like to emphasize here that based on the beam dynamics considerations, the requirement on maximum achievable gradient in 1 GeV proton accelerators for SNS or ADSS applications is modest, and typically less than 20 MV/m. A stable beam with low beam loss is the primary criteria there. Based on our detailed magneto-thermal analysis, it appears that the reactor grade material with *RRR* 100 will give similar performance as RRR 300 grade material, and may therefore be acceptable. For the proposed ISNS project at RRCAT, Indore we have performed the calculations of $Q_0$ and $B_{th}$ for the 650 MHz elliptical SRF cavity geometry described in *Ref.* 18. These calculations are presented in the Fig. 8, where the variation of $Q_o$ as a function of the applied magnetic field $B_a$ is shown for a fixed value of $\sigma_{no} \sim 2.069 \times 10^9$ $(\Omega$ m$)^{-1}$ corresponding to *RRR* ~ 100.

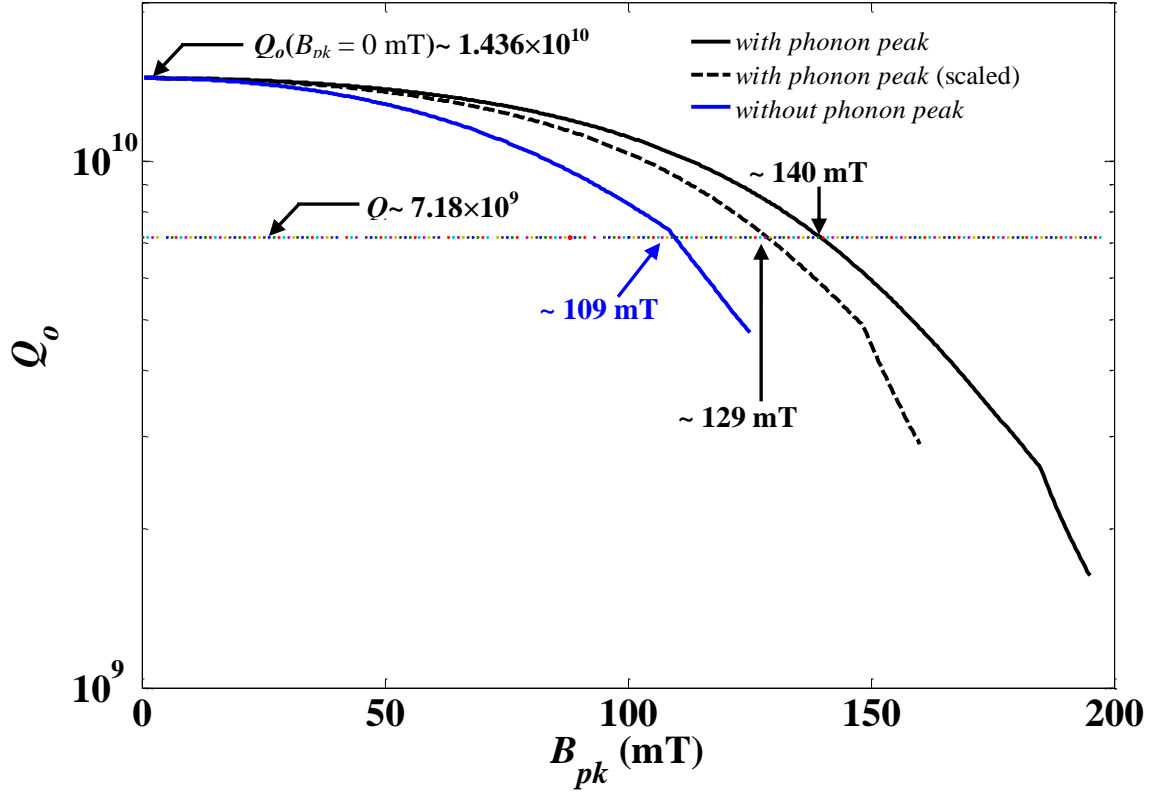

Fig. 8: Plot of $Q_0$ as a function of $B_a$, as obtained from the analysis performed on an ISNS cavity [18] for a fixed value of $RRR \sim 100$, for three possible variations of $\kappa(T)$. For these calculations, we considered 4 mm thick plate geometry. We have taken $R_i = 10$ n$\Omega$ in this analysis.

For high average power accelerator for SNS or ADSS applications, the cryogenic heat load is an important consideration. Hence, for such cases, it will be more practical to restrict the operating gradient of the cavity up to a value, where the $Q$ value drops down to not more than 50 % of the zero field $Q$ value. With these considerations, as seen from Fig. 8, the maximum magnetic field of $\sim 109$ mT can be supported at the cavity surface for the case, where we do not consider the phononic contribution. This value changes to $\sim 140$ mT, when we consider the full phononic contribution, and to $\sim 129$ mT, when we consider a scaled down phononic contribution. For these cavities the design value of $B_{pk}/E_{acc}$ is 4.56 mT/(MV/m) [19], which implies that even without considering any phononic contribution, we can go for an $E_{acc}$ of $\sim 24$ MV/m with $RRR$ 100 grade Nb, which is sufficient to fulfil the requirement. Another added advantage of using reactor grade $RRR$ 100 Nb will be that it will give nearly 45% higher value of quality factor in comparison to the cavities made of high purity (*e.g.* $RRR \sim 300$) Nb.

## IV. DISCUSSIONS AND CONCLUSIONS

In this paper, we have revisited the correlation between the purity level of the niobium SRF material, and the threshold magnetic field value $B_{th}$ for magneto-thermal breakdown of a SRF cavity. An increase in the purity level of the Nb material helps in achieving a higher thermal conductivity in its normal state. However, in the superconducting state, and in the clean limit of superconductivity (as in the case of high $RRR$ Nb materials), it is also associated with a simultaneous increase in the value of the superconducting surface resistance. Thus, this results in an ease in heat transfer, but with an added heat load. Therefore, to study the effect of material purity on the Nb-SRF cavity performance, a rigorous thermal analysis was performed for a niobium SRF cavity, considering the breakdown of the superconducting property of the material as a magneto-thermal phenomenon. In our analysis:

(1) $\sigma_{no}$ was used as a measure of the purity level of the Nb material.
(2) $R_s$ and $\kappa$ were calculated as a function of $T$, $B_a$, and the purity level of the material.
(3) Kapitza resistance was estimated as a function of $Ts$ and $T_s$.

As a first step of our analysis, we presented a case study for the 1.3 GHz TESLA cavity, considering a constant value of $\sigma_{no} \sim 2.069 \times 10^9$, which corresponds to $RRR \sim 300$ niobium material. Considering the bath temperature $T_B = 2$ K, we evaluated the maximum achievable acceleration gradient in the cavity, limited by the magneto-thermal breakdown of superconductivity. Agreement of our results with the experimentally reported observations in Ref. 1 validates our approach. After benchmarking our *magneto-thermal* analyses, we used this approach to study the influence of material purity on the performance of niobium based SRF cavity. Calculations performed without considering the phononic contribution thermal conductivity shows that for high $RRR$ grade Nb, $B_{th}$ shows a marginal increase with material purity. Interestingly, when we consider the phononic contribution that gives rise to phonon peak in thermal conductivity, $B_{th}$ reaches a maximum for modest value of $RRR$, after which it decreases and nearly saturates. We compared the $B_{th}$ for an SRF cavity made of $RRR$ 100 grade (or reactor grade) Nb with that made using RRR 300 grade Nb. Based on our magneto-thermal analysis, we find that $B_{th}$ is marginally lower for $RRR$ 100 grade compared to $RRR$ 300 grade, bur still acceptable for building 1 GeV proton/H$^-$ linac for SNS or ADSS applications, and provides nearly 45% higher value of quality factor.

We would like to mention that the results presented in this paper were obtained, considering the plate geometry of the cavity material, in order to keep the analysis simple and one dimensional. We would like to clarify that we have also repeated the calculation with a three-dimensional (3D) model of an elliptic SRF cavity half-cell in ANSYS using ANSYS® APDL, where, considering the azimuthal symmetric of the cavity, only a $15^0$ sector of the half-cell was modeled to minimize the computational effort. The field profile used in that calculation was obtained from the electromagnetic eigen-mode analysis of the cavity. The results obtained using this model were within 10 % the results obtained using the simplified plate geometry. The proximity between these two set of result establishes that due to the small thickness of the cavity wall, the heat flows effectively in one direction.

In brief, we performed a detailed magneto-thermal analysis to find an optimum value of the purity level of the material especially for the SNS or ADSS application. Our Study thus shows the required value of $RRR$ of niobium material to be used for making cavities for 1 GeV superconducting linac for SNS/ADSS applications can be reduced to 100 from 300, which is currently followed as the desired specification by the SRF community worldwide. From our literature study, we could not find a definite scientific basis behind the choice of $RRR \sim 300$. As mentioned in Ref. [30], the choice was based on the availability of pure Nb materials with a gross assumption of superior superconducting properties in such high purity materials. We thus believe that the choice of $RRR \sim 300$ is somewhat empirical, which might have been chosen under certain conditions, and then the trend was continued. There are some experimental results on SRF cavities that are made of niobium having $RRR < 300$ [31], which convey a similar point. In this context, we would like to mention a study on the systematic trends for the medium field $Q$-Slope reported in Ref. 32 by J. Vines *et al*. In their analysis too, they have performed a thermo-magnetic calculation to study the trend of threshold magnetic field for the onset of $Q$-Slope, considering few $RRR$ values. They have however not calculated the value of the threshold magnetic field, and more importantly they have only considered a partial interdependence of $R_s(T, B_a)$, $\kappa(T)$ and $h_k(T)$ while varying the purity level (*i.e. RRR*) of niobium during the numerical analysis. Nonetheless, they observed the increasing trend of the threshold magnetic field with the reduced value of RRR, which supports our analysis.

Considering from the point of view of ease of availability, *RRR* 100 or reactor grade niobium can possibly be tested for the fabrication of SRF cavities. Table-4 of ASTM B393 shows that the strength of the reactor grade niobium is 30% higher than that of *RRR* 300 grade niobium. This gives further possibility of reducing the cavity thickness, which will benefit in two ways – (*i*) reducing peak temperature at the cavity surface, thereby increasing the threshold field, and (*ii*) bringing down the material weight for each cavity.

We would like to highlight that in our analysis, we have used the normal state electrical conductivity ($\sigma_{no}$) at low temperature as a measure of the material impurity rather than *RRR*. It is to be noted that knowing the *RRR* alone does not provide us the value of the electrical conductivity at low temperature (in normal state). This is because *RRR* is the ratio of the room temperature resistivity and resistivity at low temperature, and both these quantities have dependence on the purity level of the material. In Ref. 22 (as well as several references), a constant value of room temperature resistivity ($\rho_{no\ 295\ K} \sim 14.5 \times 10^{-8}$ $\Omega$ m [24]), independent of the purity level, is assumed to extract the information about low temperature resistivity for a given value of *RRR*. This methodology is therefore not very appropriate. We therefore feel that a better approach is to directly use the low temperature normal state electrical conductivity ($\sigma_{no}$) instead of *RRR* as a measure of the purity level of the material.

At the low temperature regime, restoration of the phonon peak improves thermal conductivity $\kappa$. As observed experimentally, post-processing profoundly influences the lattice or the phonon contribution in $\kappa$. On the other hand, in the low temperature, even if the material is normal, the Wiedemann-Franz formula alone cannot predict the total electronic thermal conductivity $\kappa_{en}(T)$. An added contribution comes in the form of $aT^2$, as described by Eq. (6), which is significant at the low temperature. Although $a$ can be estimated theoretically [26], it is not in good agreement with the experimental measured value. We have used the experimentally measured value of $a$ in our analysis. As it is described in Section II, scale factor $R(y)$ scales down $\kappa_{en}(T)$ to $\kappa_{es}(T)$, when the material becomes superconductor. Therefore, instead of specifying only *RRR* for the starting Nb material, we suggest that $\kappa$ (*T*) could also be an important parameter for Nb materials specification. Note that the $\kappa$ (T) plays an important role in determining the diffusivity $\alpha$ of the material. Here, $\alpha = \kappa_{es}/\rho \times C_P(T)$, where, $\rho$ and $C_P$ [33] are the density and specific heat of the material. We thus believe that instead of specifying the *RRR*, we should specify $\sigma_{no}$, $\kappa$ and $\alpha$ of niobium to get full details of the material properties that determine the SRF cavity performance. Taking the reactor grade niobium as a material for cavity fabrication, we can specify these material parameters at 9.3 K as $\sigma_{no} \sim 6.89 \times 10^8$ $\Omega$ m$^{-1}$, and $\kappa \sim 138.68$ W m$^{-1}$ K$^{-1}$ and $\alpha \sim 0.005$ m$^2$ s$^{-1}$. Here, $C_P$ (*T* ~ 9.3 K) = 3.36 J Kg$^{-1}$ K$^{-1}$.

To further emphasize our point of view, here, we would like to mention the recent activities of nitrogen or titanium doping in niobium SRF cavities [34, 35, 36], which results in lowering the mean free path, thus reducing material purity / *RRR* at the surface, while maintaining the high value of material purity / *RRR* in the bulk. This helps in achieving the lower value of surface resistance $R_s$, while maintaining a high value of bulk thermal conductivity $\kappa$. These recent trends corroborate our finding that the lower RRR is helpful in getting better performance from niobium SRF cavities. However, more importantly, in the case of "Nb / Ti doping", one has to first produce high *RRR* grade niobium, which will be expensive, and then dope at 1400°C (for Ti doping), and at 1000°C (for N$_2$ doping), which results in the threshold magnetic field of 90 mT (for Ti doping) and 40-70 mT (for N$_2$ doping) to obtain a high $Q_o$. On the other hand, as suggested in our paper, it is economically more viable to use reactor grade niobium with low RRR, without any doping, with much higher threshold magnetic field.

In our analysis, we have considered the global breakdown phenomenon of the superconducting property of the Nb material in the context of an SRF cavity. In some cases, the local effect like crack or micro-crack on the surface, inclusion of a large bead of normal or magnetic material and/or rough welding pits/bumps may also cause hot spots, which in turn can trigger the breakdown of superconductivity of the material. Such extraneous effects can, however, be avoided by proper inspection and screening of starting Nb materials, and implementing careful SRF cavity fabrication and post processing techniques.

To conclude, we have analyzed the effect of material purity on the threshold RF magnetic field value $B_{th}$ on the cavity surface that determines the limiting acceleration gradient in a Nb-based SRF cavity. Based on our analysis, we argue that *RRR* ~ 300 grade niobium seems to be an over specification. This specification of Nb materials can be relaxed, which will have important implication in terms of a significant reduction in the cost of a Nb SRF cavity.

ACKNOWLEDGEMENTS
One of us (ARJ) would like to thank Amit Kumar Das for fruitful discussions.